\documentclass[]{aa}
\usepackage[varg]{txfonts}
\usepackage{color}
\usepackage{graphicx}
\usepackage{natbib}
\usepackage{ulem}
\usepackage{soul}
\usepackage{float}
\usepackage[breaklinks=true]{hyperref} %% to avoid \citeads line fills
\bibpunct{(}{)}{;}{a}{}{,} %% natbib format like A&A and ApJ
\newcommand{\eg}{{e.g.}}
\newcommand{\ie}{{\it i.e.}} 
\newcommand{\rs}{$R_\odot$}

\def\aap{A\&A}

\def\apjl{Astrophys. J. Lett.}

\begin{document}

\title{North-South asymmetry in the magnetic deflection of polar coronal hole jets}
\author{G. Nistic\`o\inst{1}  \and G. Zimbardo\inst{2} \and S. Patsourakos\inst{3} \and V. Bothmer\inst{4} \and V. M. Nakariakov\inst{1,5,6}}
\institute{Centre for Fusion, Space and Astrophysics, Department of Physics, University of Warwick, Coventry CV4 7AL, UK\label{1}\\
\email{g.nistico@warwick.ac.uk}
\and
{Dipartimento di Fisica, Universit\`a della Calabria, Arcavacata di Rende, 87036 (CS), Italy\label{2}}
\and{Department of Astro-Geophysics, University of Ioannina, Greece\label{3}}
\and{Institut f\"ur Astrophysik, University of G\"ottingen, Germany\label{4}}
\and{Central Astronomical Observatory at
Pulkovo of the Russian Academy of Sciences,
St Petersburg 196140, Russia\label{5}}
\and
{School of Space Research, Kyung Hee University,
Yongin, 446-701, Gyeonggi, Korea\label{6}}}

\date{Received \today /Accepted dd mm yyyy}

\abstract
{Measurements of the magnetic field in the interplanetary medium, of the sunspots area, and of the heliospheric current sheet position, reveal a possible North-South asymmetry in the magnetic field of the Sun. This asymmetry could cause the bending of the heliospheric current sheet of the order of 5--10 deg in the southward direction, and it appears to be a recurrent characteristic of the Sun during the minima of solar activity.}
{We study the North-South asymmetry as inferred from measurements of the deflection of polar coronal hole jets when they propagate throughout the corona.}
{Since the corona is an environment where the magnetic pressure is greater than the kinetic pressure ($\beta \ll 1$), we can assume that magnetic field controls the dynamics of plasma. On average, jets during their propagation follow the magnetic field lines, highlighting its local direction. The average jet deflection is studied both in the plane perpendicular to the line of sight, and, for a reduced number of jets, in three dimensional space. The observed jet deflection is studied in terms of an axisymmetric magnetic field model comprising dipole ($g_1$), quadrupole ($g_2$), and esapole ($g_3$) moments.}
{We measured the position angles at 1 \rs~ and at 2 \rs~of the 79 jets from the catalogue of \citet{Nistico09}, based on the STEREO ultraviolet and white-light coronagraph observations during the solar minimum period March 2007-April 2008. We found that the propagation is not radial, in agreement with the deflection due to magnetic field lines. 
Moreover, the amount of the deflection is different between jets over the north and those from the south pole. Comparison of jet deflections and field line tracing shows that a ratio $g_2/g_1 \simeq -0.5$ for the quadrupole and a ratio $g_3/g_1 \simeq 1.6-2.0$ for the esapole can describe the field. The presence of a non-negligible quadrupole moment confirms the North-South asymmetry of the solar magnetic field for the considered period.}
{We find that the magnetic deflection of jets is larger in the North than in the South {of the order of 25-40\%}, with an asymmetry which is consistent with a southward deflection of the heliospheric current sheet of the order of 10 deg, consistent with that inferred from other, independent, datasets and instruments.}


\titlerunning{N-S asymmetry of polar jet deflection}
\authorrunning{Nistic\`o et al.}
\maketitle
%%%%%%%%%%%%%%%%%%%%%%%%%%%%%%%%%%%%%%%%%%%%%%%%%%%%%%%%%%%%%%%%%%%%%%%%%%%%%
%% Main text
\section{Introduction}

The solar corona is an environment highly structured by the strength and the topology of magnetic
fields. Even during solar minima, the corona is far from being a quiet region but it evolves on many time scales, including the solar cycle.
The observations of the corona in EUV and X-ray wavelengths reveal bright (dense) regions, coinciding 
with the presence of active regions, and extended dark (void) areas, named coronal holes. In the same way, white-light observations obtained from coronagraphs show the presence of ray-like features at higher latitudes, and helmet streamers at middle and equatorial latitudes. 
In astrophysics, knowledge of the solar magnetic field comes mainly from measurements of splitting atomic lines due to the Zeeman effect. This method can be applied to the radiation coming from the photosphere, allowing to estimate the vector magnetic field, but there is not much possibility to have direct measurements of the magnetic field of the corona. Indirect estimates of the coronal magnetic field are obtained from extrapolations techniques (PFSS, NLFF) \citep{Wiegelmann2012}, by radiophysical methods and coronal seismology \citep[\eg][]{Nakariakov01}.

There are several observations suggesting a North-South (N-S) asymmetry of the solar magnetic field during solar minima \citep[\eg, see the discussion in][]{Erdos2010}. Early evidences came from direct measurements of the photospheric magnetic field by magnetograms, which were extrapolated to the solar wind source surface by \citet{Hoeksema95}, who showed that the magnetic field strength in the Sun's south polar cap was 60\% larger than in the north one during some solar minima.
Then, the Ulysses spacecraft provided new insights of the N-S asymmetry: thanks to its orbit, nearly perpendicular to the ecliptic plane, Ulysses explored high latitudes regions of the heliosphere, and during its passage close to the Sun, the so-called fast latitude scan at about 1.4 AU, gave us measurements of the 
interplanetary magnetic field \citep{Erdos98} and particle data. Indeed, an indication of N-S asymmetry comes from the global distribution of the solar wind speed \citep{Tokumaru2015}, the latitudinal gradients of energetic particle fluxes \citep{Simpson96,Heber96b} which show an unbalance of fluxes betweeen north and south of $6\%-15\%$. Moreover, \citet{Erdos2010} studied and compared magnetic field data of Ulysses during the first latitude scan (coincident with the minimum of solar cycle 22 in 1994-95) and the third one (minimum of solar cycle 23 in 2007-08), and shows that the radial component of the magnetic field (normalised to 1 AU) of the South hemisphere is greater by a factor 1.12--1.21 than that of the North. In addition they found a southward shift of the heliospheric current sheet (HCS) of the order of 3--5 deg \citep[see also][]{Virtanen10}. A previous limit of about 7 deg was found by \citet{Mursula04} based on heliospheric magnetic field observations at 1 AU. As of now, the actual values of the southward shift of the heliospheric current sheet remain not well known.

A possible model for explaining the N-S asymmetry is to consider the global magnetic field as the contribution of several multipole components, and ascribe the mismatch between the north and the south magnetic fields to the contribution of the quadrupole moment, as suggested by \citet{Bravo00} and \citet{Mursula04}. Indeed, Fig. 1 of \citet{Bravo00} and \citet{Bravo01}, which compares the dipole, quadrupole and esapole structures, shows how this asymmetry could be generated: if we look at the direction of magnetic field for the dipole and the quadrupole during two consecutive solar minima, we note that they are opposite in the north pole, and concordant in the south pole. The direction of the magnetic field of the esapole is always concordant with that of the dipole at the poles. This means that, due to the action of the quadrupole moment, the northern magnetic field is weakened, while the southern field is enhanced, causing the evident asymmetry and the shift of the heliospheric current sheet in the southward direction.

Observations of non-radial white-light coronal streamers \citep{Wang1996} or deflection of EUV polar plumes \citep{dePatoul2013a,dePatoul2013b} allowed to study the possible configuration of the global and polar magnetic field, respectively. On the other hand, coronal jets are transient density enhancements of plasma that can highlight the local structure of the magnetic field. Indeed, hot jets at poles observed in EUV or X-rays are naturally explained as the result of magnetic reconnection between emerging flux with the open magnetic field: the tension force of the disconnected field lines compresses the plasma and pushes it away along the magnetic field lines \citep{Yokoyama1995}.

The aim of this work is to estimate the configuration of the solar magnetic field, and consequently the N-S asymmetry, starting from the latitudinal deflection of polar jets observed by STEREO spacecraft. Section 2 presents the analysis of the deflection of the polar coronal jets from the catalogue of \citet{Nistico09}, which regards jets observed from March 2007 to April 2008, which roughly corresponds to the solar minimum, between the end of solar cycle 23 and the beginning of 24. The analysis is performed in 2D and 3D space. Modeling and extrapolation of the coronal magnetic field structure is explained in Section 3, and conclusions are given in Section 4. 

\section{Polar jet deflection measurements}

 \subsection{2D analysis}
\begin{figure*}[htbp]
\begin{center}
\includegraphics[width=16 cm]{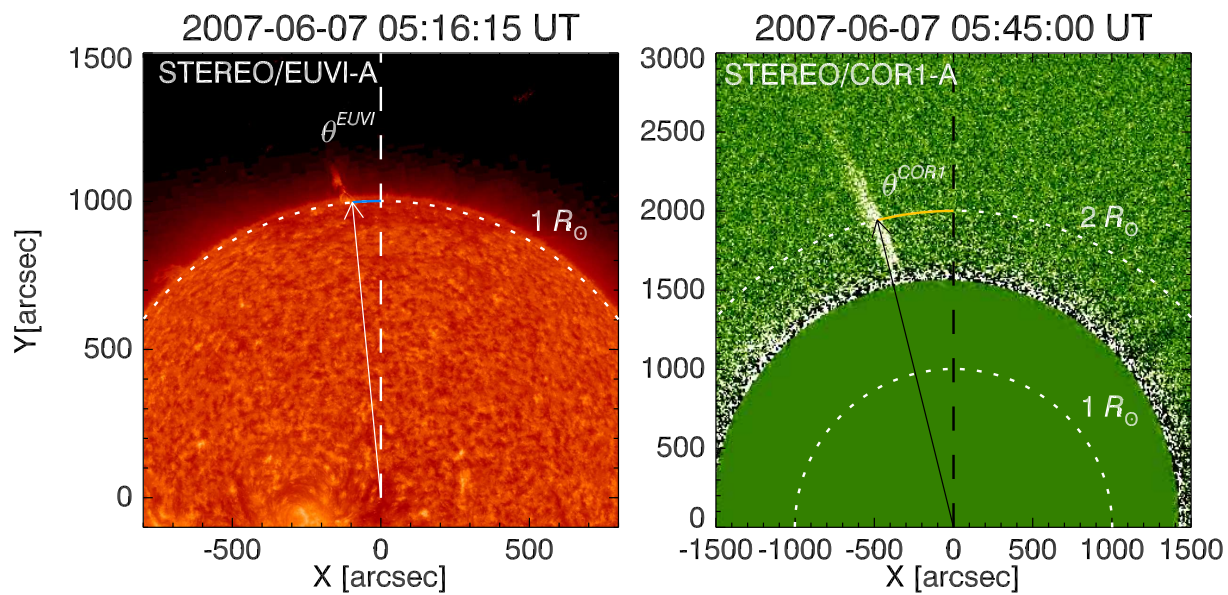}
\end{center}
\caption{Image of the Sun at 304 \AA~(left) from  EUVI and from COR1 in white light (right). PAs are from the north solar axis in the counterclockwise direction (see the reference frame overlapped on the images) at 1 \rs~in the EUVI FOV and at 2 \rs~in the COR1 FOV.}
\label{fig4_02}
\end{figure*}

\begin{figure}[htbp]
%\begin{tabular}{p{4cm} p{4cm}}
%\includegraphics[width=4.5 cm]{figure/fign.eps} &
%\includegraphics[width=4.5 cm]{figure/figs.eps} \\
\centering
\includegraphics[width=9 cm]{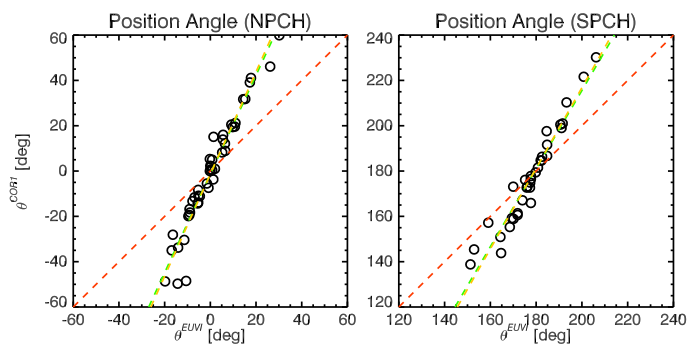}
\caption{Plots of the EUVI PAs (horizontal axis) $vs$ the COR1 PAs (vertical axis) for jets occurring at the north (left) and at the nouth polar coronal hole (right). The red dashed line represents the bisector of the plane and ideally should trace equal PAs between the EUVI and COR1 FOV (radial propagation); the green dashed line fits the data points. Events rooted significantly away from the poles in EUV tend to have an even larger deviation in COR1.}
\label{fig4_03}
\end{figure}

We investigate the influence of the large scale coronal magnetic field on the motion of 79 jets through the corona observed with STEREO during March 2007-April 2008, which are catalogued in \citet{Nistico09}. During this period, the angular separation between the two STEREO spacecraft increased from 2 to $\sim$48 deg.
We exploited data from EUVI and COR1 instruments of the SECCHI package.
Our method consists of calculating the position angle (PA), \ie~the angular displacement from the north axis to the jet position in counterclockwise direction, at two fixed distances from the solar centre: 1 \rs, that corresponds to the solar limb, and 2 \rs~(see Fig. \ref{fig4_02}). 
Since the EUVI images cover the full disk up to 1.4 \rs, they are appropriate for measuring the PA at 1 \rs~($\theta^{EUVI}$), that 
in most cases corresponds with the position of the jet footpoint. COR1 images have a FOV within 1.4--4 \rs, but the exact 
limits vary due to the offset of the occulter with respect to the Sun center. In order to clearly identify the jet in COR1, we took PAs at 2 \rs~($\theta^{COR1}$), somewhat above the boundary of the occulter from the Sun centre.
% We used FITS file data {\bf and measure the PA for each single jet} from {\bf either} STEREO A or B, according to {whichspacecraft gives a better view} where the jet is better seen. 
FITS file data were prepped within the SolarSoft (SSW) environment by \texttt{secchi\_prep.pro}, which allows to calibrate and apply corrections to the images, including rotation of the axis towards the solar North. Information about the Sun's centre in units of pixels are retrieved for EUVI from the keywords CRPIX1 and CRPIX2 of the header, for COR1 we used the function \texttt{wcs\_get\_pixel.pro}\footnote{The keywords CRPIX1/2 for COR1 return the centre of the occulter. Coordinates information can be retrieved with the Word Coordinate System (WCS) routines in SSW starting from \texttt{fitshead2wcs.pro}. See \url{http://hesperia.gsfc.nasa.gov/ssw/gen/idl/wcs/wcs_tutorial.pdf}.}, the pixel size in arcsec is given from CDELT1/2, and the solar radius from RSUN. Circumferences defining the distance of 1 and 2 \rs~from the centre are over-plotted in the images and the PA for each single jet is visually determined and measured in deg either from STEREO A or B, according to which spacecraft offers a better view in both instruments. We assume an error of $\Delta l=$10 pixels in the location of the jet (this apparently large error can include some other instrumental errors, jitter, etc.) with both instruments. This can be easily converted in terms of angular displacement, by considering the pixel size of each detector ($\Delta_{pix}$) and the radial distance from the centre, which are not constant through the period of observations. Thus, a typical error in EUVI is $\Delta \theta^{EUVI} = \Delta l~\Delta_{pix}/r \approx 1 \deg$, ( with $\Delta_{pix}\approx1.6''$ and $r \approx 950''$), and in COR1 $\Delta \theta^{COR1} \approx2 \deg$ (with $\Delta_{pix} \approx 7.5''$ and $r \approx 1900''$).
%%%% FIG 1 %%%%%%%%
%%%%%%%%%%%%%%%%%%%%%%%%%%%%%%%%%%%%%%%%%%%%%%%
The measurements of the PAs for the 79 jets from COR1 are plotted as a function of PAs from EUVI in Fig. \ref{fig4_03} for the North pole (left) and the South one (right).  
%%%%%%%%%%%%%%%%%%%%%%%%%%%%%%%%%%
%%%%%%% FIGURE 2%%%%%%%%%%%%%%%%%%% 
%%%%%%%%%%%%%%%%%%%%%%%%%%%%%%%%%%%%%%%%%%%%%%%
The somewhat larger number of events in the north coronal hole with respect to the south (45 against 34, respectively) is due to the orbital features of STEREO, which allow a better view of the northern region in the investigated time period \citep{Nistico09}. In addition, during the period of observations the Sun was at minimum. The jet angular positions, represented by circles, are fitted by a linear function separately for the north/south pole coronal hole (NPCH/SPCH), according to the equation:
\begin{equation}
\theta^{COR1}= a~\theta^{EUVI}+ b
\label{eq4_1}
\end{equation}
where $a$ is the slope of the line and $b$ is the constant term of the linear fit. We performed linear fits with two different IDL routines: the first is a fit unweighted by measurement errors with \texttt{linfit.pro}, which performs regression of the data points along the vertical distance (dashed green lines in Fig. \ref{fig4_03}), and the second with \texttt{linfitex.pro} of the MPFIT package 
\citep{Markwardt2009}, which performs full-Cartesian regression by taking into accounts errors in both $x$ and $y$ variables (yellow line). Both the fits are almost coincident. 
The red line is a reference line corresponding to the case when jets had the same PAs at 1 and at 2 \rs, implying radial propagation.
Although some scatter of the data points is present, an overall trend is evident:
jets having small PAs (if in the NPCH) or small displacements from the south polar axis (if in the SPCH) in EUVI FOV show a small 
deviation in the COR1 FOV (they are near or on the red line, representing events that have the same PA when seen from both instruments); jets having large PAs (if in the NPCH) or large displacement from the south polar axis (if in the SPCH) in the EUVI FOV (i.e., events which occur at lower latitudes) show greater deviation in COR1 on average. This can be associated with the fact that the trajectory of the jets is not simply radial but bends towards the equator: the actual jet angle is the difference between the position vectors at the heights of 1 and 2 \rs (see the green dashed line that does not coincide with the red one). We can assume that jets propagate, on average, along the magnetic field lines, which are almost radial near the solar dipole axis, while those at lower latitudes deviate more markedly from the radial direction because of the dipolar structure.  
This is also consistent with the over-expansion towards low latitudes of the fast solar wind in polar coronal holes \citep[\eg,][]{Fisk96}. 
The non-radial outward propagation of the jets is a property consistent with those of other coronal structure: \eg, coronal streamers and polar plumes extend non-radially \citep{Wang1996}, as well as non-radial seems to be the propagation of CMEs, as found in earlier studies \citep[\eg,][]{Cremades04}.
It can be noticed that when going from the EUVI to COR1, the changes in PA are
larger in the north coronal hole than in the south one. More precisely, linear fits of data points, with \texttt{linfit.pro} and \texttt{linfitex.pro} for the North and the South give the values of the parameter $a$ and $b$ that are summarized in Table \ref{tab4_1}. 
\begin{table*}[htpb]
\begin{center}
\caption{Value{\bf s} of parameters for the linear fits of the EUV and COR1 PAs, latitudes, and longitudes for jets at the North and South poles.}
%Values of $a$ and $b$ from the fits of the EUVI and COR1 PAs of the analysed jets, respectively for the North and the South pole. The {\bf fits} of the latitudes and longitudes from 3D analysis are shown in the table as well. The bending factors from the fits of the PAs and latitudes are indicated as $a_N$ for the north, and $a_S$ for the south. The value of the quantities $(a_N-a_S)/a_S$ and $(a_N/a_S)^2$ is also listed.}
%\caption{Values of the parameters $a$ and $b$ (see Eq. \eqref{eq4_1}) resulting from the fits of the EUVI and COR1 PAs of the analysed jets, respectively for the North and the South pole in 2D using \texttt{linfit} and \texttt{linfitex} procedures. The {\bf fits} of the latitudes and longitudes from 3D analysis are shown in the table as well. The slopes for the linear fits of the PA and latitudes are indicated as $a_N$ for the norther jets, and $a_s$ for the souther ones. The quantity $(a_N-a_S)/a_S$ shows that the slope $a_N$ is around 22--43\% larger than $a_S$. Notice the square of the ratio $a_N/a_S$ of the fitted slopes is also listed. The physical meaning of this quantity is discussed in the last section.}
\label{tab4_1}
   \begin{tabular}{l | c c c c c c}
   \hline
   \hline
              &    $a_N$  & $b_N$ & $a_S$ & $b_S$  & $(a_N-a_S)/a_S$  &  $(a_N/a_S)^2$ \\
             &                  &  [deg]  &             &  [deg]   &                                   &                  \\    
             \hline
{\bf PA}        &                  &            &             &             &                                   &                   \\             
LINFIT         & $2.18 \pm 0.09$ & $-1.29 \pm 0.97$ & $1.72 \pm 0.09$ & $-129.20\pm 15.44$ & 27\% &  $1.60 \pm 0.21$ \\
LINFITEX    & $2.25 \pm 0.04$ & $-1.31 \pm 0.45$ & $1.78 \pm 0.04$ & $-138.93 \pm 6.98$ & 26\% & $1.61 \pm 0.09$ \\ 
\hline
{\bf Latitude}&                 &            &             &              &                            &                  \\
LINFIT         & $1.77 \pm 0.36$ & $ -72.33\pm 28.59$ & $1.45 \pm 0.24$ & $  47.64\pm 18.26$ &  22\% & $1.49 \pm 0.78$ \\   
LINFITEX    &  $2.31 \pm 0.06$ & $-115.51 \pm 4.97$ & $1.62 \pm 0.04$ & $56.74 \pm 3.06$ & 43\% & $2.04 \pm 0.15$ \\
\hline
{\bf Longitude} &            &             &            &              &                            &                   \\
LINFIT        & $1.11 \pm 0.07$ & $ -10.06\pm  4.66$ & $1.00 \pm 0.02$ & $  -1.53\pm  2.42$ &  -  &  -  \\
LINFITEX   & $1.04 \pm 0.01$ & $2.08 \pm 0.51$ & $0.98 \pm 0.01$ & $0.81 \pm 0.77$ & -  & -  \\         
\hline
\hline
\end{tabular}

\end{center}
\end{table*} 

\begin{figure}[htbp]
\begin{center}
\includegraphics[width=9 cm]{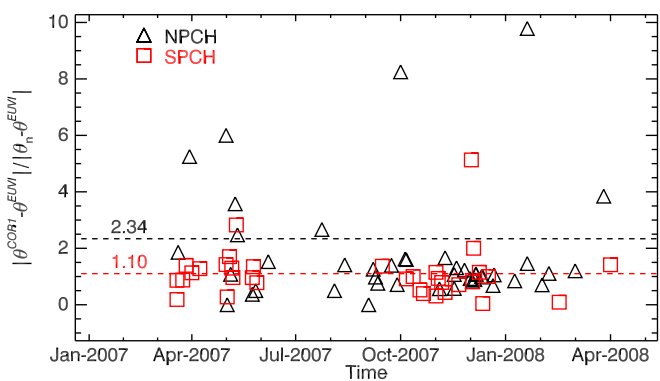} \\
\caption{Absolute angular displacement normalized to the EUVI PA relative to the solar axis as a function of time for jets occurred in the North (hollow black triangles) and in the South (hollow red squares). The quantity $\theta_n$ is 0 deg for events at the North and 180 deg for those at the South, in order to compare values between jets seen at the two poles. The dashed horizontal lines mark the average displacement: 2.34 deg for jets at the North and 1.10 deg for those at the South.}
\end{center}
\label{fig4_04}
\end{figure}

The parameter $a$, which represents the slope of the fitting line for the PAs, and hence a measure of the average bending of jets, is indicated as $a_N$ for the north (second column), and $a_S$ for the south (forth column) in Table \ref{tab4_1}. We also list the values of the parameter $b$, as $b_N$ for the North (third column) and $b_S$ for the South (fifth column), although they will not have a relevant role in the discussion. The bending is found to be on average greater in the North than in the South, giving an indication that jets are more deflected in the North than in the South ($a_N/a_S > 1$). The value of the quantity $(a_N-a_S)/a_S$ shows that $a_N$ is around 26\% larger than $a_S$. The square of the ratio of the coefficients, $(a_N/a_S)^2$, is also listed. The physical meaning of this quantity will be discussed in the last section.
This asymmetry is also evident in Fig. \ref{fig4_04}, where we plot the absolute value of the relative jet-bending from the solar axis $|\theta^{COR1}-\theta^{EUVI}|/|\theta_n-\theta^{EUVI}|$ as function of time for north polar jets (black void triangles) and south polar jets (red void squares). The parameter $\theta_n$, which represents the PA of the solar axis, is 0 deg at the North and 180 deg at the South. We can infer that there is not a particular temporal dependence of the PA displacement from EUVI to COR1 FOV on time, as might be the case for Ulysses measurements during the fast latitude scans, and the average magnetic deflection is larger at the North pole than at the South pole.

%%%%%%% FIGURE 3 %%%%%%%%%%%%%%%%%%%%%%%%%%%%%
%%%%%%%%%%%%%%%%%%%%%%%%%%%%%%%%%%%%%%%%%%%%%%%

The evidence that jets are more deflected in the north pole than in the south pole could be associated with the different bending of the open magnetic field lines: in our case, jets occurring in the North, are more deviated from their origin since the magnetic field lines are more curved than in the South during this period of observations.

\subsection{3D analysis}
 
The distinctive capability of STEREO is to perform 3D stereoscopic geometry or trajectory reconstruction of solar structures, such as loops \citep[\eg][]{Nistico13}, and CMEs \citep{Bosman2012}. Understanding the 3D evolution of jets through the corona as observed with the EUVI and COR1 instruments can provide additional information about their magnetic deflection. 
In order to examine this aspect, we measured the 3D position of jets in Stonyhurst longitude and latitude, at given radial distances of 1 and 2 \rs.
In practise, we used the routine \texttt{scc\_measure.pro}, available from the SSW package, which allows to determine these quantities. We did this for a subset of the jets in the catalogue, i.e., for those events for which the determination of the 3D position is more reliable because of better viewing; this was possible for 38 events (20 at the North and 18 at the South, respectively). We identified the coordinates at 1 \rs~by triangulating the base of the jet in EUVI images. For each event we collected 10 measurements, in order to take into account errors due to the triangulation process, and calculated the average values and standard deviations for the longitude and latitude. The same procedure has been performed for COR1, in order to measure coordinates at 2 \rs. In this case, collecting points at this fixed distance required more efforts, since there is not any reference that can help us in locating the jet (in the EUVI FOV, the footpoint of the jet or the limb is a good marker). The visibility of the jet, in both cases, has been eventually improved by using difference images. 

The results of the 3D measurement are shown in the top panels of Fig. \ref{fig4_05}.
The jet location is de-projected in polar plots, showing the Stonyhurst longitude (concentric circles) and latitude (radial lines) at 1 \rs~as red dots, and at 2 \rs~as green dots. A blue dashed line connects the locations of a jet, which is marked by a number according the catalogue from \citet{Nistico09}. The radial lines and arcs centered on the dots are the error bars for the latitude and the longitude, respectively. It is worth noting that these graphs provide an anticipation of what we can see with Solar Orbiter, when it will be able to see directly the polar cap when orbiting out of the ecliptic plane. The first impression is the consistency of the measurements from EUVI and COR 1, which are taken independently, showing displacements toward low latitudes at higher distance from the Sun, and almost a radial trajectory (i.e. a small longitudinal shift).  
In a similar way as done for the PAs, we can fit the measurements between 1 and 2 \rs~for the latitudes and longitudes, respectively, by using \eqref{eq4_1}. The scatter plots are shown in Fig. \ref{fig4_05}. The latitudes for both poles (top panels) are distributed far away from the bisector of the plane (dashed red line, which ideally should mark events with no change in latitudes), and the linear fits (in green with \texttt{linfit} and yellow with \texttt{linfitex}) return a slope $a_N= 1.78-2.31$ for the North, and $a_S=1.44-1.62$ for the South (see the second row group in Table \ref{tab4_1}). In this case, the quantity $(a_N-a_S)/a_S$ shows that $a_N$ is around 22--43\% larger than $a_S$, which is almost consistent with the result{\bf s} obtained from the PA measurements. The ratio $(a_N/a_S)^2$ is around 1.5--2.0. On the contrary, the longitudes (third row group in Table \ref{tab4_1}) are very close to the bisector (slopes are 1.11--1.04 and 1.00-0.98 for the north and south jets, respectively), suggesting that the studied jets did not exhibit a significant shift in the azimuthal direction. 
%%% FIGURE 4 %%%%%%%%%%%%%%%%
\begin{figure*}[htbp]
\begin{center}
   \includegraphics[width=18 cm]{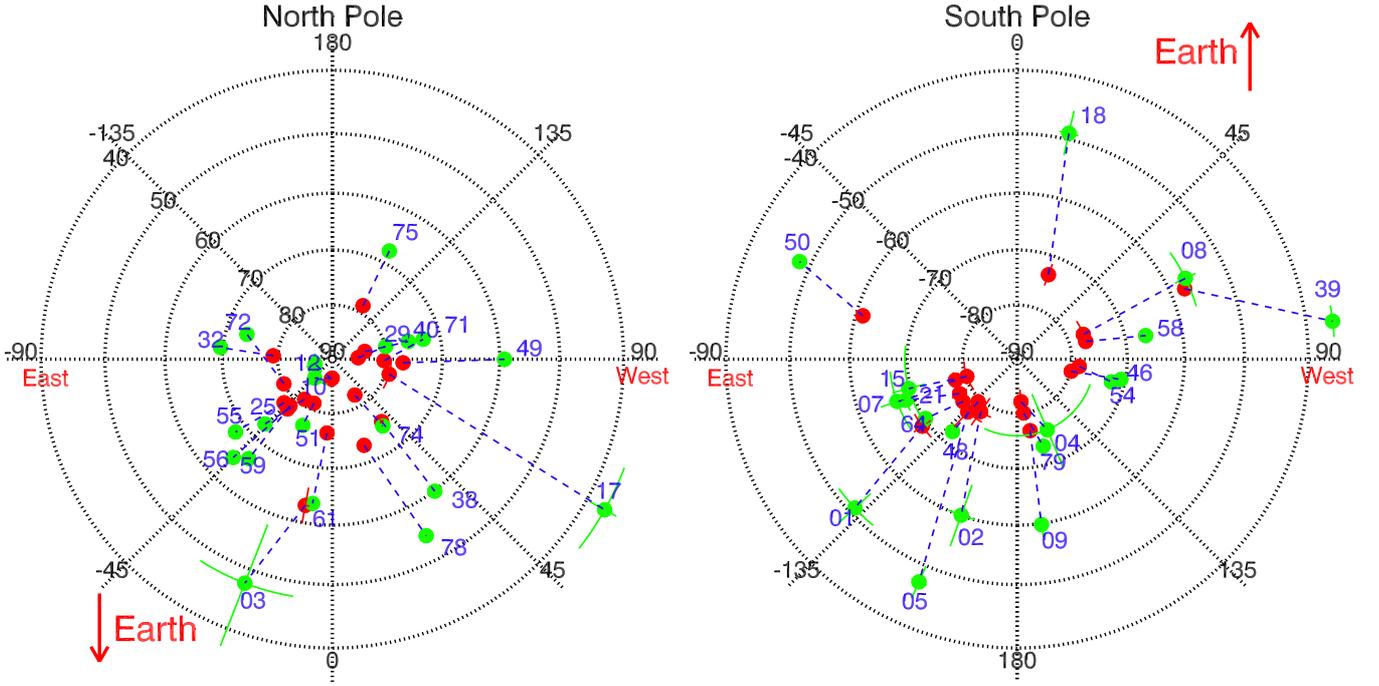}
 \end{center}
 \caption{Polar plots of 3D positions of jets seen at the North (left) and South poles (right). The circles represent the heliographic longitudes, and the radial lines the latitude meridians measured in deg. Position of jets at 1 R$_\odot$ are in red, at 2  R$_\odot$ in green, respectively. The Sun-Earth direction is given by a red arrow in both plots.}
 \label{fig4_05a}
\end{figure*}
 
%%% FIGURE 5 %%%%%%%%
 \begin{figure*}
  \begin{center}
  \sidecaption
   \includegraphics[width=12 cm]{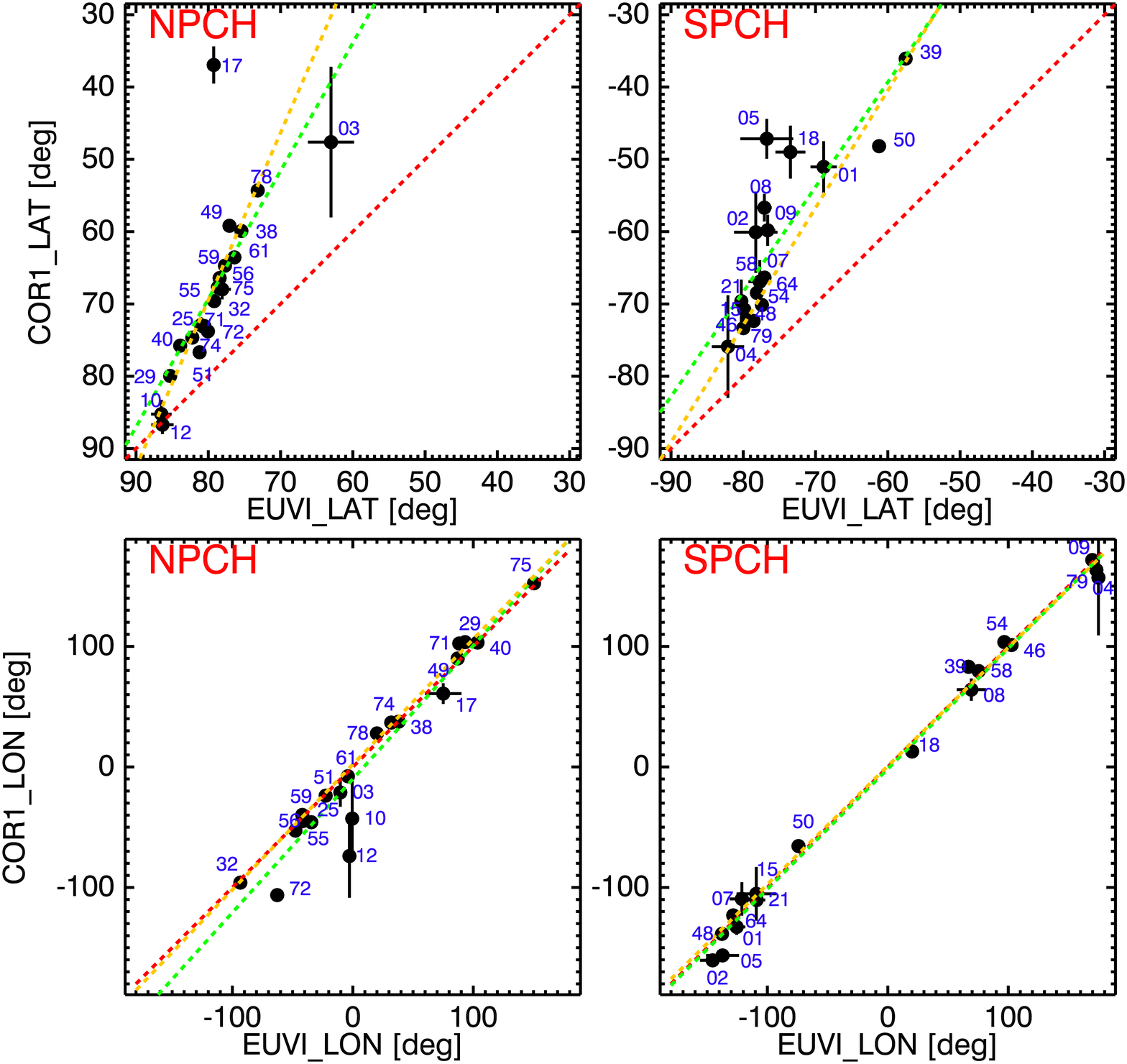}
\caption{Top: scatter plots of the latitudes measured with EUVI at 1 \rs, and COR1 at 2 \rs~for the north (left) and south (right) polar jets. Linear fits of the points is given by the green and yellow dashed lines, while the red one is the bisector of the plane. Bottom: similarly to the previous graphs, scatter plots of the longitudes. The numbers in blue identify the events in the catalog of \citet{Nistico09}.}
\label{fig4_05}
\end{center}
\end{figure*}
%%%%%%%%%%%%%%%%%%%%%%%%%%
\section{A model for the coronal magnetic field} 

From the results shown in the previous section, a natural question arises: ``Is the different deflection of jets, in the North and in the South poles, an indication of a magnetic North-South asymmetry?''.
There are several extrapolation methods for estimating the magnetic configuration of the solar corona \cite[for details see,][chapter 5]{Altschuler69,Kivelson95,Aschwanden05}. 
In the potential field (\ie, current-free) approximation, the magnetic field can be obtained as the gradient of a scalar potential, ${\bf B} = -\nabla \Psi$, and the scalar potential function can be expressed in spherical coordinates as an expansion in terms of the Legendre polynomials $P^m_l(\cos\theta)$:   
\begin{equation}
\Psi(r,\theta, \phi)=R_\odot \sum_{l=1}^N \sum_{m=0}^l f_l(r) P_l^m(\cos{\theta})[g_l^m \cos{(m\phi)}+h_l^m \sin{(m\phi)}].
\label{eq4_3}
\end{equation}
The indices $l$ and $m$ are integer numbers: $l$ is strictly positive and defines the number of axes of symmetry of the field, while $m$ can assume $2l+1$ values $(-l,...,-1,0,1,...,l)$ and defines the orientation of the axes in spherical geometry. The function:
\begin{equation}
f_l(r)=\frac{\left( r_w/r \right)^{l+1} -\left( r/r_w \right)^{l}}{\left( r_w/R_\odot \right)^{l+1} - \left( R_\odot/r_w \right)^{l}},
\label{eq4_4}
\end{equation}
 fixes the position of the solar wind source surface at $r_w$ solar radii; beyond $r_w$ the magnetic field lines are purely radial, reproducing the configuration of the magnetic field in the solar wind \citep{Aschwanden05}.
Thus, Eqs. \ref{eq4_3}-\ref{eq4_4} must be used only for  $R_\odot < r < r_w$.

The components of the  magnetic field can be found as the derivative of the scalar potential $\Phi$.
%\begin{eqnarray}
%B_r(r,\theta, \phi) & = & -\frac{\partial \Phi}{\partial r},\\ 
%\label{eq4_5}
%B_\theta(r,\theta,\phi) & = & -\frac{1}{r} \frac{\partial \Phi}{\partial \theta},\\
%\label{eq4_6}
%B_\phi(r,\theta,\phi) & = & -\frac{1}{r\sin{\theta}}\frac{\partial \Phi}{\partial \phi}. 
%\label{eq4_7}
%\end{eqnarray}
We can further simplify the expressions of the magnetic field components, if we assume axial symmetry. This assumption is supported by the 3D analysis of the jet position, showing negligible shift in the longitudinal or azimuthal direction: then the component $B_\phi$ is null. Of course, this assumption implies that we are neglecting other effects like the possible magnetic dipole tilt. Axial symmetry is enforced by setting $m=0$. Indeed, near the poles ($\theta \simeq 90\deg,180\deg$), all associated Legendre polynomials $P_l^m \propto \sin^m\theta \to 0$, except for $m=0$.
%;  so that, the magnetic field components can be written as:
%\begin{eqnarray}
%B_r(r, \theta) & \simeq & -R_\odot \sum_{l=1}^N \frac{\partial}{\partial r} f_l(r) P_l^0(\cos{\theta}) g_l^0 , \\
%\label{eq4_8}
%B_\theta(r,\theta) & \simeq &-\frac{R_\odot}{r} \sum_{l=1}^N f_l(r) \frac{\partial}{\partial \theta} P_l^0(\cos{\theta}) g_l^0 , \\
%\label{eq4_9}
%B_\phi & = & 0 \, . 
%\label{eq4_10}
%\end{eqnarray}

The expansion now depends only on the index $l$. If we truncate the series at $l=3$, we have three contributions that give the dipole ($l=1$), the quadrupole ($l=2$), and the esapole (at $l=3$) terms,
with the corresponding moments (or harmonic coefficients) $g_1, g_2, g_3$ (we dropped the superscript $m$ in the coefficients since it is always 0).% The contribution of the monopole ($l=0$) is not taken under consideration, since it is physically unacceptable.

After some algebra, we find the magnetic field components due to the dipole:
\begin{eqnarray}
B_r^{(1)}(r,\theta) & = & \left( \frac{R_\odot}{r}\right)^3 \left( \frac{2r_w^3+r^3}{r_w^3-R_\odot^3}\right) g_1  \cos{\theta}, \\
\label{eq4_11}  
B_\theta^{(1)}(r,\theta) & = & \left( \frac{R_\odot}{r}\right)^3 \left( \frac{r_w^3-r^3}{r_w^3-R_\odot^3}\right) g_1  \sin{\theta};  
\label{eq4_12}
\end{eqnarray}
the components due to the quadrupole:
\begin{eqnarray}
B_r^{(2)}(r,\theta) & = & \frac{1}{2}\left( \frac{R_\odot}{r}\right)^4 \left( \frac{3r_w^5+2r^5}{r_w^5-R_\odot^5}\right) g_2  (3\cos^2{\theta}-1), \\
\label{eq4_13}
B_\theta^{(2)}(r,\theta) & = & 3\left( \frac{R_\odot}{r}\right)^4 \left( \frac{r_w^5-r^5}{r_w^5-R_\odot^5}\right) g_2  \cos{\theta}\sin{\theta};
\label{eq4_14}
\end{eqnarray}
and the components due to the esapole:
\begin{eqnarray}
B_r^{(3)}(r,\theta) & = & \frac{1}{2}\left( \frac{R_\odot}{r}\right)^5 \left( \frac{4r_w^7+3r^7}{r_w^7-R_\odot^7}\right) g_3  (5\cos^3{\theta}-3 \cos{\theta}), \\
\label{eq4_15}
B_\theta^{(3)}(r,\theta) & = & \frac{1}{2}\left( \frac{R_\odot}{r}\right)^5 \left( \frac{r_w^7-r^7}{r_w^7-R_\odot^7}\right) g_3  (15\cos^2{\theta}\sin{\theta}-3\sin{\theta}).
\label{eq4_16}
\end{eqnarray}

In the limit of $r_w \rightarrow \infty $, we have the classical expressions for the dipole, the quadrupole, and the esapole in free space. 
The resulting magnetic field can be written as the sum of the dipole, quadrupole, and esapole contributions:
\begin{eqnarray}
B_r(r, \theta) & = & B_r^{(1)}+B_r^{(2)}+B_r^{(3)}, \\
\label{eq4_23}
B_\theta(r,\theta) & = & B_\theta^{(1)}+B_\theta^{(2)}+B_\theta^{(3)}.
\label{eq4_24}
\end{eqnarray}

The magnetic field lines can be obtained by integrating, with a Runge-Kutta scheme of the 4th order for example, the two first order differential equations:
\begin{equation}
\left\{
  \begin{array}{c c c}
   dr/ds & = & B_r/B, \\
         &   &               \\
   d\theta/ds & = & B_\theta/rB. \\
  \end{array}   
\right.
\label{eq4_33}
\end{equation}
where $s$ is the distance along the field line. 
This model is then used to find a relationship between the coefficients $a$ and the normalised multipole coefficients of the reduced field model.

\subsection{Magnetic moments from the Wilcox Solar Observatory}

Given the model for the coronal magnetic field with the assumption of axial symmetry and the truncation to the dipole, quadrupole and esapole moments, we can search which values of the magnetic moments $g_1$, $g_2$, $g_3$, are suitable for better describing the jet magnetic deflections.  For comparison, these coefficients are calculated from magnetograms data, provided and published by the Wilcox Solar Observatory (WSO) (see the website \url{http://wso.stanford.edu/}). 
For inferring values of these coefficients from magnetograms, two kind of hypotheses are made regarding the inner boundary conditions in the photosphere \citep{Wang92}. Indeed, from magnetograms we can measure the line-of-sight (LOS) component of the photospheric field at a given latitude $\alpha$ on the solar disk. The {\it ``classic''} model takes into account  the projection of the LOS photospheric field along the radial and latitudinal component ($B_r=B_{LOS} \cos \alpha, B_\theta=B_{LOS} \sin \alpha$, with $\alpha$ the line-of-sight angle); conversely, the {\it ``radial''} model assumes that the photospheric field is totally radial in the photosphere $(B_r=B_{LOS}/\cos \alpha, B_\theta=0)$. 
Since the magnetic field is non potential and nearly radial at the photosphere \citep{Wang92},  a better approach is  considered to be the radial model. 

\begin{figure*}[htpb]
  \centering
 \includegraphics[width=16 cm]{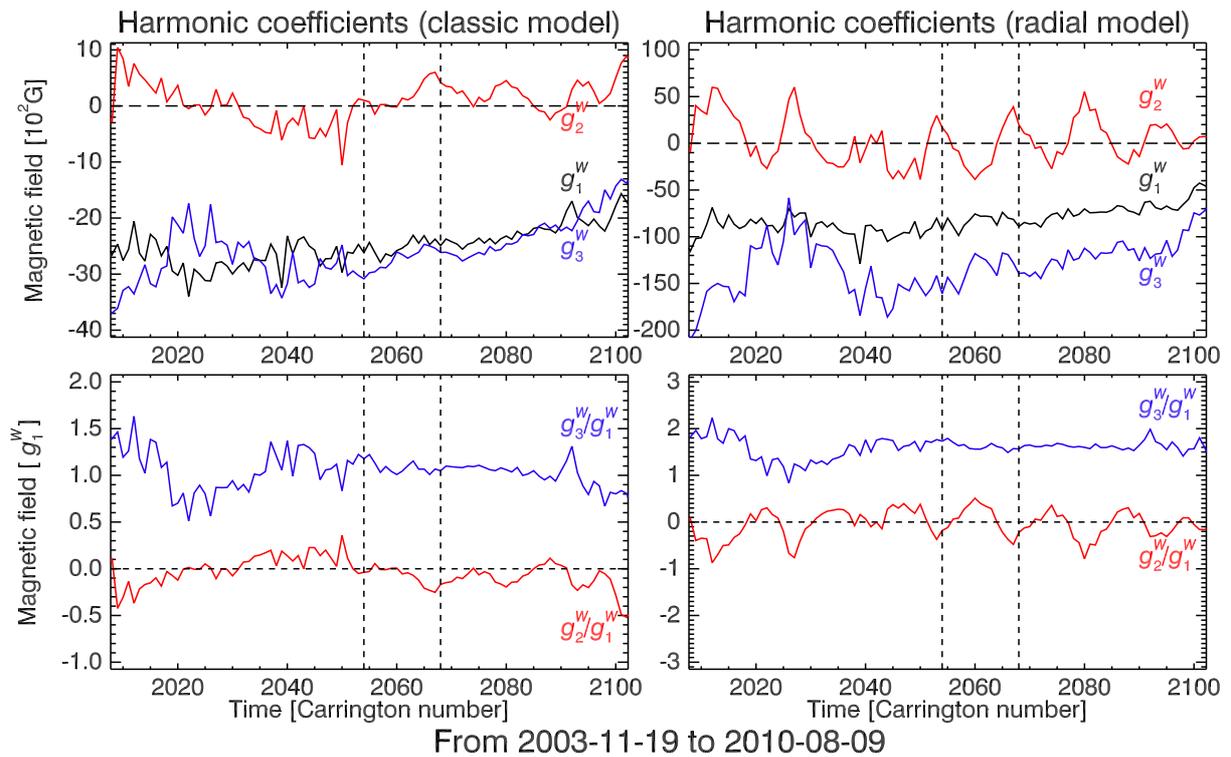} 
  \caption{Top: Time evolution of the harmonic coefficients $g_1^W, g_2^W, g_3^W $, as calculated by the Wilcox Solar Observatory (\url{http://wso.stanford.edu}) in the classic (left) and radial (right) model. Bottom: time evolution of the quadrupole and esapole moments normalised to the dipole component. Note the periodic change of the quadrupole over a scale of 20 Carrington rotation ($\sim$2 yr) in the radial model. The vertical dashed lines enclose the period of our observations.}
\label{fig4_2}
\end{figure*}

The top panels of Fig. \ref{fig4_2} show the temporal evolution of the coefficients for the ``classic'' and the ``radial'' model, as calculated by the Wilcox Solar Observatory.
The bottom plots are the temporal evolution of the ratios $g_2^W/g_1^W$ and $g_3^W/g_1^W$ (where the superscript $W$ signifies $g$ coefficients calculated by the WSO). The region bounded by the dashed lines is the temporal window in which our jets are observed.

%%%% FIGURE 6 %%%%%%%%%%%%%%
We can see that the harmonic coefficients evolve in time, and that the quadrupole moment is less in magnitude than the other ones and exhibits an oscillatory behaviour, especially in the radial model, with alternating sign and with a period of approximately two years, indicating a possible association with the biennal oscillations \citep{Vecchio2008,Vecchio2012,Bazilevskaya2014}. On the contrary, the dipole and esapole moments show a constant sign.  Below, we try to calculate these coefficients in a different way and to compare them with those obtained from the WSO. Further, we find the associated coronal magnetic field structure and the position of the heliospheric current sheet projected on the solar surface.   
 
\subsection{Fitting the magnetic field model to the polar jets PAs}
\begin{figure}
\centering
 \includegraphics[width=9 cm]{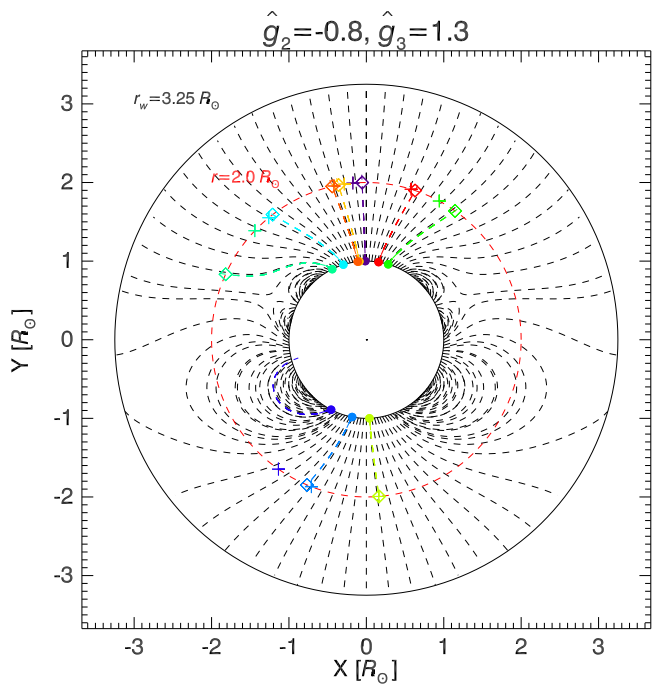}
 \caption{Example of magnetic field line integration for some jets with moments $\hat{g}_2=-0.8$ and $\hat{g}_3=1.3$. The starting points of the integration are at 1 \rs~from EUVI observations and represented as coloured dots. The integrations is made up to 2 \rs and the results (void squares ) are compared with the corresponding jet positions as observed with COR1 (plus signs). For some events, especially those at lower latitudes, the integration can result in a closed line (purple dashed line). }
 \label{fig6}
\end{figure}

Here, we estimate the best values of the coefficients $g_2$ and $g_3$ that fit our jet observations {and measurements of PA}, in terms of the dipole moment (which is used as a normalization factor). For several values of $\hat{g}_2 = g_2/g_1$ and $\hat{g}_3 = g_3/g_1$ coefficients, we integrate numerically the equations \eqref{eq4_33} from the base of the jets, \ie the PA as measured in the EUVI FOV, until 2 \rs. This yields the difference between the final PA from the numerical integration, $\theta^{COR1_{Mod}} $, which is a function of $\hat{g}_2$ and $\hat{g}_3$,  and that one measured in the COR1 FOV, $\theta^{COR1_{Obs}}$. Then we calculate the standard deviation $\sigma(\hat{g}_2,\hat{g}_3)$ as: 
\begin{equation}
\sigma(\hat{g}_2,\hat{g}_3)=\sqrt{\frac{\sum_{i=1}^{N} \left[\theta_i^{COR1_{Mod}}(\hat{g}_2,\hat{g}_3)-\theta_i^{COR1_{Obs}}\right]^2}{N-1}},
\label{sigma_def}
\end{equation}
with $N$ the total number of magnetic field lines successfully integrated from the jet base at 1 \rs~up to 2 \rs. The number $N$ is not necessarily 79 but can be less, depending on the values of the magnetic moments since the magnetic field lines can be closed without reaching 2 \rs. Fig. \ref{fig6} shows for example some magnetic field lines integrated for a few jets with given values of the coefficients $\hat{g}_2$ and $\hat{g}_3$. We can notice that in some events a good agreement between the final position from the integration and the observed PA at 2 \rs~is found, some others show a considerable gap, whilst for an event the integrated line results to be closed and does not reach 2 \rs.
%%%%%% FIGURE %%%%%%%%%%%%%%
%%%%%%%%%%%%%%%%%%%%%%%%%%
The coefficient $\hat{g}_2$ spans from -3 to 3 and $\hat{g}_3$ from 0 to 3, both in steps of 0.1.
We give the results of $\sigma(\hat{g}_2,\hat{g}_3)$ as 2D contour maps representing the value of the standard deviation as a function of the quadrupole (vertical axis) and esapole (horizontal axis) moments, using different models of the magnetic field, \ie, classic and radial, and different distances of the solar source surface $r_w=$ 2.5 and 3.25 \rs~\citep{Altschuler69,Hoeksema95}.      
The grid in the maps is determined by the varying values of $\hat{g}_2$ and $\hat{g}_3$, with a resolution of 0.1 for both parameters, as used in the numerical model.
%%%% FIGURE 7 %%%%%%%%%%%%%%%%
\begin{figure}[htpb]
 \centering
   \includegraphics[width=8 cm]{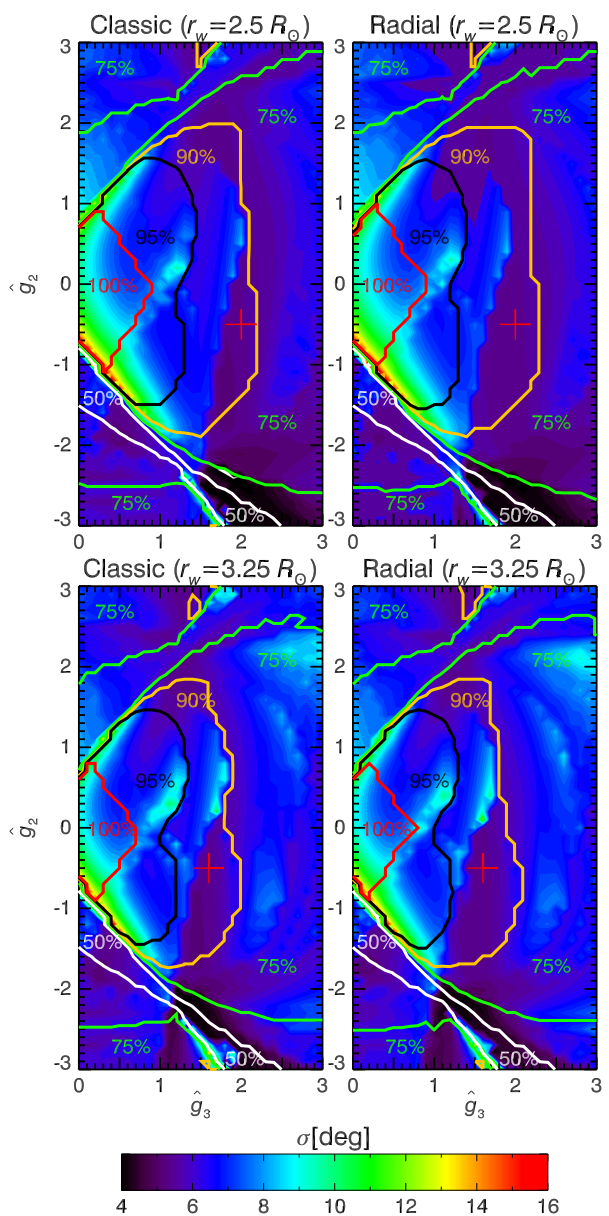} \\
 \caption{Contour maps representing the standard deviation ($\sigma$) from \eqref{sigma_def} for different values of  $r_w$ (top: 2.5 \rs, bottom: 3.25 \rs), for the classic (left) and radial (right) model, as a function of $\hat{g}_2$ (vertical axis) and $\hat{g}_3$ (horizontal axis) coefficients. The colorbar defines the variation range of $\sigma$. In addition, coloured lines enclose regions with different percentage (100, 95, 90, ...) of jets, hence magnetic field lines, which are successfully integrated. The red crosses locate the chosen minimum in a low-$\sigma$ region with at least 90\% of jets integrated. This minimum is found for $\langle\hat{g}_2\rangle = -0.5$, and $\langle\hat{g}_3\rangle= 1.6$ (bottom panels)- 2.0 (top panels).}
\label{fig4_3b}
 
\end{figure}

The top  panels of Fig. \ref{fig4_3b} show the standard deviation maps for the classic (left) and radial (right) model considering the solar source surface at 2.5 \rs; the bottom row gives the same results for $r_w= 3.25$ \rs. The colorbar provides the variation range of $\sigma$: from 4 (black) to 16 deg (red). The purpose of these maps is to give us an indication of which values of $\hat{g}_2$ and $\hat{g}_3$ minimize the standard deviation (dark regions in the maps), and better fit the deflections of jets. In addition, they are overlaid by some coloured lines that enclose portions of the maps characterised by the same percentage of jets, hence magnetic field lines, successfully integrated. Indeed,  if a certain value of $\sigma$ is obtained for different values of the magnetic moments, the number of integrated lines $N$ will provide a further constrain to discriminate which values of $\hat{g}_2$ and $\hat{g}_3$ are more reliable: the higher $N$, the better the adaptation of the magnetic field model to our jet observations. For this reason, we can exclude the minimum in $\sigma$ found for $\hat{g}_2 \sim -2.5$ and $\hat{g}_3 \sim 1.8$, since it is obtained for as few as almost 50\% of the observed jets.
 
We do not obtain specific and exclusive values of magnetic moments that fit our observations. The shape of the low-$\sigma$ is almost similar in all cases and appears to be more sensitive to the esapole moment, since it is more narrowed for some values of $\hat{g}_3$, and spread for several values of $\hat{g}_2$, but a patch with $\sigma \le 6$ deg and in the limit of $N=90\%$  is found for negative values of the quadrupole, in agreement with the values from WSO. This is marked by red crosses in the maps at values of $\langle\hat{g}_2\rangle \sim -0.5$ and $\langle\hat{g}_3\rangle \sim 1.6-2.0 $.

\begin{figure}[htpb]
 \begin{center}
    \includegraphics[width=9.0 cm]{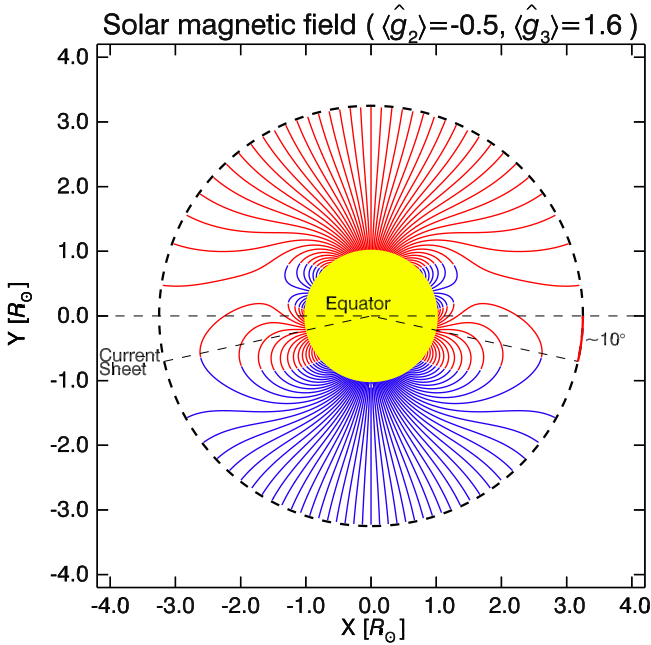}
 \end{center}
  \caption{Structure of the coronal magnetic field lines with $\langle\hat{g}_2\rangle=-0.5$ and $\langle\hat{g}_3\rangle=1.6$, and the corresponding HCS position projected on the source surface at 3.25 \rs~for the interest period of observations between March 2007 and April 2008. Inward magnetic field lines are plotted in red, outward in blu.}
\label{fig4_5b}

\end{figure}

We can now infer the structure of the coronal magnetic field by computing Eqs. (4--9) with the obtained values of $\langle\hat{g}_2\rangle$ and $\langle\hat{g}_3\rangle$ and plotting the magnetic field lines. This is shown in Fig. \ref{fig4_5b}: the HCS in the considered period of observations results to be coned southward, forming an angle of about $10 \deg$, which is broadly consistent, although somewhat larger, with some estimates found in the literature ranging between 3 and 10 deg \citep{Simpson96,Heber96b,Mursula04,Erdos2010}.    
On the other hand, the HCS is not a stationary feature and the tilt is subject to a change over the time due to the evolution of the magnetic field structure. For the period under examination, the average values of the quadrupole and esapole components, as determined by the Wilcox Solar Observatory,  are of the order of $g_2^W/g_1^W\sim-0.1,0.05$ and $g_3^W/g_1^W\sim1.0,1.6$ for the classic and radial model, respectively. In addition the quadrupole component shows a large variability, assuming positive and negative values with peaks at $\pm 0.5 g_1^W$ in the case of the radial approximation. %Estimates of the average tilt per Carrington rotation of the HCS, based upon an expansion of the magnetic field for several $l$ and  $m$ harmonic coefficients, both for the classic and radial model,  are listed at (\url{http://wso.stanford.edu/Tilts.html)}. Values of the tilt angle show a considerable spread with an average extent of the order of 15 (radial), 30 (classic model) deg, with a consistent extent both to the North (20 deg for the radial model, 30 deg for the classic one) and to the South (30 for the radial model, and 40 deg for the classic).
%%%% FIGURE 8 %%%%%%%%%%%%

\section{Discussion and conclusions}

In this work, we used polar corona jets as a probe for understanding the magnetic field structure of the solar corona during a solar minimum. 
Since the corona is an environment with a plasma-$\beta$ parameter less than 1, we assumed that jets on average propagate along the magnetic field lines. For simplicity, we also assumed that the large scale solar magnetic field during a solar minimum is axisymmetric. 
We measured the PA of jets at 1 \rs~in the EUVI FOV, and at 2 \rs~in the COR1 FOV, and analysed the deflection of jets. We found that jets are deflected toward low latitudes, in agreement with the fact that the magnetic field lines are bent toward the equator, and this deflection is more pronounced in the North pole than in the South pole. This North-South asymmetry has been found in other datasets, starting from photospheric magnetic field measurements \citep{Hoeksema95}, global distribution of the solar wind speed \citep{Tokumaru2015}, analysis of latitudinal gradient in solar energetic particles \citep{Simpson96,Heber96b}, and also magnetic field measurements in the interplanetary medium by the Ulysses spacecraft \citep{Erdos2010}.
This asymmetry can be modelled in terms of multipole components of the global magnetic field \citep{Bravo00,Mursula04}: during solar minima the quadrupole moment tends to influence the total magnetic field, weakening it in one pole and enhancing it in the opposite one.
We have estimated the contribution of the quadrupole moment, starting from jet PA data, and comparing our results with those of the WSO.
We expressed the coronal magnetic field as the sum of the dipole, quadrupole, and esapole moments, starting from the expression of a scalar potential function $\Phi$ in terms of spherical harmonic expansion. We traced magnetic field lines from the footpoint of jets (at 1 \rs) until 2 \rs~for different values of the magnetic moments. We obtained the harmonic coefficients $\hat{g}_2$ and $\hat{g}_3$, normalised to the dipole, that minimize the standard deviation of position angle differences coming from the numerical simulation and the observations. We obtained as reliable values $\langle\hat{g}_2\rangle=-0.5$ and $\langle\hat{g}_3\rangle=1.6$. From these estimates, we were able to compute the structure of the magnetic field lines (Fig. \ref{fig4_5b}) in which the heliospheric current sheet (HCS) shows an offset of about 10 deg, a value consistent with the results obtained by \citet{Mursula04}, but somewhat larger than that found by \citet{Erdos2010}, whose observations however correspond in part to different periods and are taken at much larger distances from the Sun, around 1.4 AU during the fast latitude scans of Ulysses.  

\begin{figure}[htpb]
    \begin{center}
     \includegraphics[width=4.0 cm]{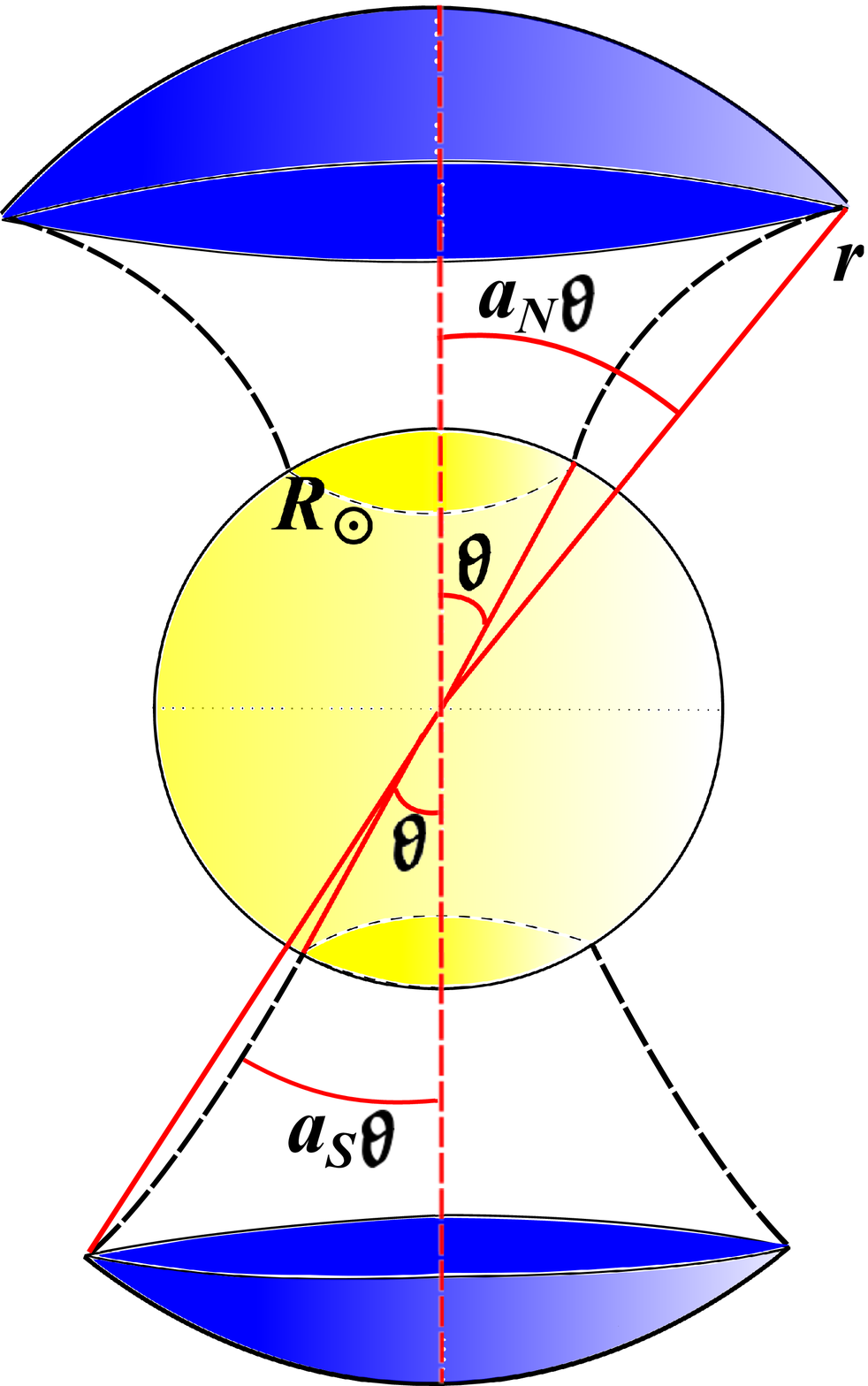}
    \end{center}
     \caption{Sketch representing the N-S asymmetry of the solar magnetic field in terms of magnetic fluxes through the polar caps. The polar caps on the solar disk at the distance of 1 \rs~represent ideally regions in polar coronal holes with equal area, which depend upon the half-opening angle $\theta$. Because of the asymmetric over-expansion of the magnetic field lines between the two hemispheres (represented by the curved dashed lines departing from the solar disk), the angle $\theta$ at a given distance $r$ will increase of a factor $a_N$ in the North and $a_S$ in the South, with $a_N>a_S$. The resulting size of the Northern polar cap will be larger than the Southern one.}
     \label{fig_sketch}
  \end{figure}

In our analysis, we show that the slopes of the linear fits for the PA and the latitudes (from 3D measurements) are different between the two poles. The angular coefficients of the fits can be immediately related to the ratio of the magnetic field values $B_S/B_N=1.12-1.21$, as reported in Table 1 from \citet{Erdos2010}, according to the following geometric interpretation in terms of conservation of magnetic fluxes through the Sun's poles.
Consider the sketch of the Sun in Fig. \ref{fig_sketch}, with the polar caps marked by dashed lines at the distance of 1 \rs, which can essentially represent areas embedded in open field regions, such as the polar coronal holes. We can express the magnetic fluxes $\Phi$ approximately as the product of the area $A$ of the polar cap with an average polar magnetic field $B$. The surface of a spherical cap depends on the half opening angle $\theta$ as $A=2\pi R_\odot^2 (1-\cos\theta)$, which we assume equal for both poles at $1$ \rs, for example. The angle $\theta$ is analogous to the PA in our measurements. Hence, because of the asymmetry, the flux $\Phi_S$ in the South pole will be larger than $\Phi_N$ in the North.  Thus, we have at the distance of 1 \rs:
 \begin{equation}
	 A_N(R_\odot)=A_S(R_\odot) ~\Rightarrow~ \frac{\Phi_N(R_\odot)}{\Phi_S(R_\odot)} = \frac{B_N(R_\odot)}{B_S(R_\odot)}
	 \label{eq_flux1}
 \end{equation}
 
Moving away of the Sun, the magnetic field  diminishes with the distance $r$ and the area of the projected polar cap must increase because of the flux conservation ($\Phi_i(R_\odot)=\Phi_i(r)$ with $i=N,S$). We can consider the variation of the magnetic field normalised to the distance $R_\odot$ as $B_i(r)=\hat{B}_i(r) B_i(R_\odot)$, and the expansion of the polar caps with the distance can be addressed in terms of a variation of the opening angle $\theta$ of a factor $a$, which is different between the two hemispheres (and also depends on the radial distance). Thus, at a certain distance $r$ (which is taken to be $R_\odot < r \le r_w$ in the PFSS model) the polar cap areas can be expressed as:
 \begin{equation}
   A_i(r) = 2\pi r^2 \left[1-\cos\left(a_i(r)\theta\right) \right] 
 \end{equation}  
 with $i=N,S$. A second order approximation for the cosine function gives $\cos(a\theta) \approx 1-\frac{1}{2}a^2\theta^2$, and, finally, the area can be expressed as $A_i(r)=\pi r^2(a_i^2\theta^2)$. By taking into account the ratio of the magnetic fluxes between North and South at $r$, we have:
 \begin{equation}
   \frac{\Phi_N(r)}{\Phi_S(r)} = \frac{B_N(r) A_N(r)}{B_S(r) A_S(r)} \approx \frac{\hat{B}_N(r)}{\hat{B}_S(r)} \frac{ B_N(R_\odot) }{B_S(R_\odot)} \left(\frac{a_N(r)}{a_S(r)}\right)^2,
   \label{eq_flux2}
 \end{equation}

and by combining eqq. \eqref{eq_flux1} and \eqref{eq_flux2} in virtue of the magnetic flux conservation, we obtain the final relation:
\begin{equation}
      \frac{\hat{B}_S(r)}{\hat{B}_N(r)} \approx \left(\frac{a_N(r)}{a_S(r)}\right)^2,	
\end{equation}
which links the estimated deflections with the ratio of the magnetic field magnitudes.
 The squared values for the ratio $a_N/a_S$ at $r=2$ \rs, according our analysis of the jet deviations, are around 1.5--2.0 (see Table \ref{tab4_1}), almost 25--{\bf 65}$\%$ larger than the ratio of the magnetic field estimated by \citet{Erdos2010}, in agreement with the larger estimate of the HCS offset in our analysis. The agreement between our results and those from \citet{Erdos2010} are rather satisfactory. We would like to outline the comparison is made on the basis of different observable (jet deflection against interplanetary magnetic field measurements normalised to 1 AU) and refer in part to different periods on time. In addition, a list from the Wilcox Solar Observatory of the tilt of the HCS reports values greater than 10 deg  for the period under interest.
%%%%% FIGURE 9 %%%%%%%%%%%%%%%%

Therefore, we have an independent indication that the solar magnetic field can indeed exhibit a N-S asymmetry, a result that can have profound implications on the models of solar dynamo.
The forthcoming missions Solar Probe Plus and Solar Orbiter will have a crucial role in assembling more precisely the extent of the N-S asymmetry, thanks to both in situ measurements of the nearly coronal magnetic field and remote observations. In particular, the Solar Orbiter UV instruments will allow to accurately check the size of the polar coronal holes and the deflection of polar jets, allowing to better constrain the solar magnetic field.

\begin{acknowledgements}
We would like to thank the referee, Bernd Inhester,  and the Editor, Hardi Peter,  for their useful and  fruitful comments.
Data are courtesy of the STEREO/SECCHI team and the Wilcox Solar Observatory. G.N. and V.M.N. thank support from STFC consolidated grant ST/L000733/1. S.P acknowledges support from the European Union (European Social Fund -- ESF) and Greek national funds through the Operation Program ``Education and Lifelong Learning'' of the national Strategic Reference Framework (NSRF) - Research Funding Program: Thales. Investing in knowledge society through the European Social Fund, as well as from an FP7 Marie Curie Grant (FP7-PEOPLE-2010-RG/268288). V.B. acknowledges support from the German Space agency DLR through the project CGAUSS. CGAUSS (Coronagraphic German And US Solar Probe Plus Survey, Grant 50 OL1201) is the German contribution to the WISPR camera currently under development for the NASA SPP mission. V.M.N. work is further supported by the European Research Council under the SeismoSun Research Project No. 321141, and the BK21 plus program through the National Research Foundation funded by the Ministry of Education of Korea.
\end{acknowledgements}

% for the bibliography, at the end
\bibliographystyle{aa} % style aa.bst
\bibliography{references} % your references Yourfile.bib
\end{document}